\documentclass[12pt]{article}
\usepackage[cp1251]{inputenc} % inputenc всегда включается ДО babel

\usepackage{graphicx}
\title{Spiral galaxies with non-typical mass-to-light ratios. }
\date{}
\author{A.S. Saburova, E.S. Shaldenkova,  A.V. Zasov}

\begin{document}
\maketitle

\begin{abstract}
Total mass-to-light ratio $M/L_B$  for $S0-Irr$ galaxies, where $M$
is the dynamical mass within the optical radius $R_{25}$, increases
systematically with $(B-V)_0$ color, but slower than that is
predicted by stellar population evolution models. It shows that the
mean ratio between dark halo and stellar masses is higher for more
``blue'' galaxies. However some galaxies don't follow this general
trend. The properties of galaxies with extremely high and extremely
low values of $M/L_B$ ratios are compared, and different factors,
accounting for the extremes, are analyzed. The conclusion is that in
some cases too high or too low $M/L_B$ ratios are associated with
observational errors, in other cases - with non-typical dark halo
mass fraction, or with peculiarities of disc stellar population.
Particularly, discs of some galaxies with low $M/L_B$ ratios turn
out to be unusually ``light'' for their luminosity and colors, which
indicates a substantial deficit of low mass stars as the most
probable cause of low $M/L_B$.
\end{abstract}

\maketitle

\section{Introduction.}

Total mass-to-light ratio $M/L$, just as mass-to-light ratio $M_*/L$
for a stellar disc is an important parameter that reflects both dark
matter (dark halo) contribution to the total mass of a galaxy and
the properties of the disc stellar population (the age distribution
of stars, stellar initial mass function $IMF$, and to a lesser
degree - the stellar population metallicity). Direct measurements
show that in most cases the total mass-to-light ratios does not fall
outside the limits in the range of 1-10 in the $B, V, R$ bands (here
and below the solar units of $M_\odot /L_\odot$ are used) and are
close to unit in near infrared bands. Hence, $M/L$ ratios which
differ significantly from their typical values may indicate either
peculiar stellar population, or extremely low (high) relative mass
of dark matter (dark halo), or may be the result of errors of
luminosity and/or mass estimations.

Luminosity of stellar population $L$ is estimated more reliably than
total mass of a galaxy. The main systematic error of luminosity
estimation comes from the uncertainty of corrections for the
internal dust extinction, which is significant for highly inclined
galaxies. Photometric parameters, reduced to face-on orientation of
galaxies  by applying the
 statistically determined corrections, are available in different
catalogs and databases (see, for example, NASA
 Extragalactic Database, \cite{ned}, HyperLeda, \cite{leda}).

The situation with the measurement of total mass of galaxies appears
to be more complicated. The direct mass measurements can be done
only on the basis of stellar or gas dynamics, but even in this case
the results depend strongly on the assumed model of a galaxy,
especially if we are interested to find the  masses of its main
components: the  disc, the bulge and the dark halo. As far as the
concept of a total mass of a galaxy is uncertain enough, the mass
estimations are often related to some fixed radius. It's convenient
to use the radius $R_{25}$, corresponding to the isophote
$25^m/sq.arcsec$ in the $B$ band, as the boundary one. This radius
contains nearly total luminosity of a disc, unless low surface
brightness (LSB) galaxies are considered.

The halo mass is often comparable to mass of a stellar disc within
the optical radius of a galaxy, although the question of
universality of the ratio of these masses for galaxies of a given
type remains opened. In general, the higher is the dark mass
fraction and the lower is the fraction of young massive stars, the
higher is the $M/L$ ratio for the galaxy. To separate these two
factors the additional information is needed, such as a color of the
disc or shape of rotation curve. Therefore it's of interest to
compare the galaxies with extremely high and low total $M/L$ ratios
to reveal the most probable reason of their extreme $M/L$ values.

\section{Total $M/L$ ratios of disc galaxies.}

To determine the influence of the dark halo and stellar population
on the total  $M/L$ ratio, the optimum way is comparison of
dynamical and photometric estimations of masses. Both methods are
model dependent. Photometric method of stellar mass measurement is
based on stellar population evolution models, which predict a strong
correlation between broad-band  colors and $M_*/L$ ratios of stellar
component. This correlation slightly depends on both star formation
history and the effect of selective absorption, (see for example
Bell, de Jong, \cite{Bell}, Portinari, \cite{Portinari}).
Theoretical evaluations of $M_*/L_B$ or $M_*/L_V$, obtained for
models with slowly varying (for example exponential declining in
time) star formation rate (SFR) and for usually accepted IMFs, lead
to values of about one solar unit for the bluest galaxies with
active star formation and 5-10 for ``red'' galaxies consisting of
old stars.

In practice, in some cases the mass values, obtained by photometric
and dynamical methods give strongly different results, although for
high luminous galaxies with accurate rotation curves the results are
usually in a reasonable agreement (see Salucci al., \cite{Salucci},
Kassin et al., \cite{Kassin},  de Blok et al., \cite{Blok}).
Nevertheless, the obtained correlation between colors and $M_*/L$
ratios, where $M_* $ is found from the rotation curve modeling, is
not so tight as it is theoretically expected (see, for example, the
diagrams, presented by Bell, de Jong, \cite{Bell}, Graham,
\cite{Graham}, Giraud et al., \cite{Giraud}, McGaugh,
\cite{McGaugh}, Barnes et al., \cite{Barnes}, Yoshino, Ichikawa,
\cite{Yoshino}). One may suppose, that a loose correlation is
associated not only with a low accuracy of mass determination, but
also with a bad choice of parameters of photometric models, such as
a stellar IMF, to the real galaxies.

The estimation of the total dynamical mass of a galaxy is more
reliable, than that for the disc. The mass within $R_{25}$ can be
obtained with the reasonable accuracy  from the equation:
\begin{equation}
\label{formula1} M=K \cdot V^2R_{25} / G
\end{equation}
where $K\approx 1$ is a coefficient, which depends on mass
distribution within the galaxy, and $V$ is a circular velocity at
$R_{25}$.  As the first approximation it may be assumed, that
$K~=~1$ and $V= W_{HI}$/$2\sin i$, where $W_{HI}$ is HI line width,
and  $i$ is the disc inclination. The $M/ L_B $ ratios found by this
way usually lay in the range between several solar units up to 10 -
20 units. High $M/ L_B $ ratios ($M/ L_B
> 10$) are very rare and occur mostly in dwarfs or in LSB-galaxies.
Indeed, there are only two galaxies with $M /L_B
> 10$ (UGC 3303 ($M / L_B = 31$) and NGC 5128 ($M / L_B = 17$)) among
the nearly a hundred galaxies brighter than
$M_B= -16^m$ (in other words, that are not extremely low luminous
dwarfs), which are given in Karachentsev catalog of nearby galaxies
\cite{Karachentsev}. Note that for one of these galaxies, peculiar
galaxy NGC 5128 (type $S0$), a more accurate model gives lower value
of $M / L_B$, namely: 3.9 for a central part of the galaxy and about
10 within R = 25 kpc (Xui et al., \cite{Xui}). Another examples of
non-dwarf galaxies with high $M / L_B$ ratios mentioned in the
literature, are Scd-galaxy UGC 7170, $M / L_B$ of which is about 43
within the radius of 100'' (that is little bit more than $R_{25}$)
(Cox et al., \cite{Cox}), and LSB galaxy UGC 128, (Zavala et al.,
\cite{Zavala}) with $M / L_B$ = 34 within 5 disc radial scalelengths
(adopted to Hubble constant $H_0=$75~km/s/Mpc, assumed in present
work). It is worth noting that the discs of all these galaxies are
observed edge-on, hence their high $M / L_B$ may be a result of
underestimation of light extinction in these cases.

Extremely low mass-to-light ratios ($M / L_B <1$) are observed
mostly in dwarf gas-rich galaxies with active star formation.
Really, there is no any object with $M / L_B<3$ among 79 galaxies of
different types, luminosity and surface brightness, considered by
Zavala et al., \cite{Zavala}, (where dynamic masses were evaluated
within five radial scalelengths of a disc, that is little larger,
than $R_{25}$). Extremely low $M / L_B$ are expected to be found  in
starburst galaxies where young stellar population gives the main
contribution to luminosity. In this case such a galaxy should also
have a low $(B-V)$ color. The presence of massive dark halo may
nevertheless increase its ratio $M / L_B$.

The connection between the dark halo mass fraction and other
properties of galaxies is a matter of discussion. As it is reported
by some authors,  mass fraction of dark halo tends to be higher in
the galaxies with lower luminosity (mass) (Persic, Salucci,
\cite{Persic}, Yegorova, Salucci, \cite{Yegorova}, Pizagno et al.,
\cite{Pizagno}, de Blok et al., \cite{Blok}) or central surface
brightness (Zavala et al., \cite{Zavala}). Nevertheless, total
$M/L_B$ ratios within optical radius correlate slightly or do not
correlate at all neither with luminosity, nor with the mean surface
brightness or morphological type, as it is revealed by nearby
galaxies (Karachentsev et al., \cite{Karachentsev}).

Diagram ``$M$/$L_B$ -- $(B-V)_0$'', where the mass $M$ is determined
from the equation (\ref{formula1}), and the total corrected  colors
of galaxies are taken from  HyperLeda \cite{leda}, is shown in
Fig.1a. The sample consists of about 1300 S0 - Irr galaxies,
selected from HyperLeda, that are brighter than $B =14^m$, and have
the inclination $i >40^0$ (to minimize the uncertainty in $i$).
Dwarf galaxies with the luminosity $L_B < 3\cdot10^8L_\odot$ are
practically absent in the sample.

\begin{figure} [h!]
\includegraphics[width=12cm,keepaspectratio]{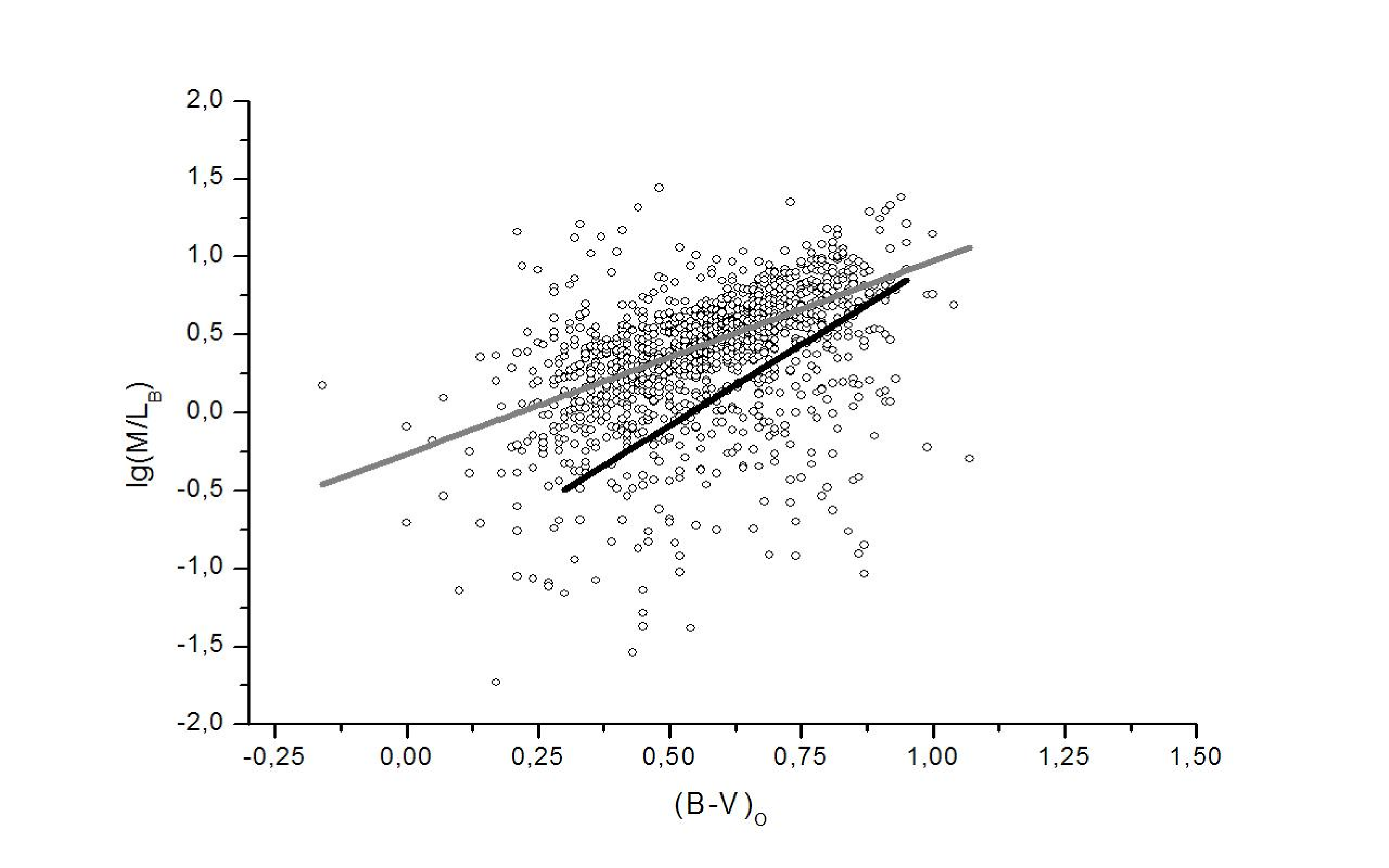}
\includegraphics[width=12cm,keepaspectratio]{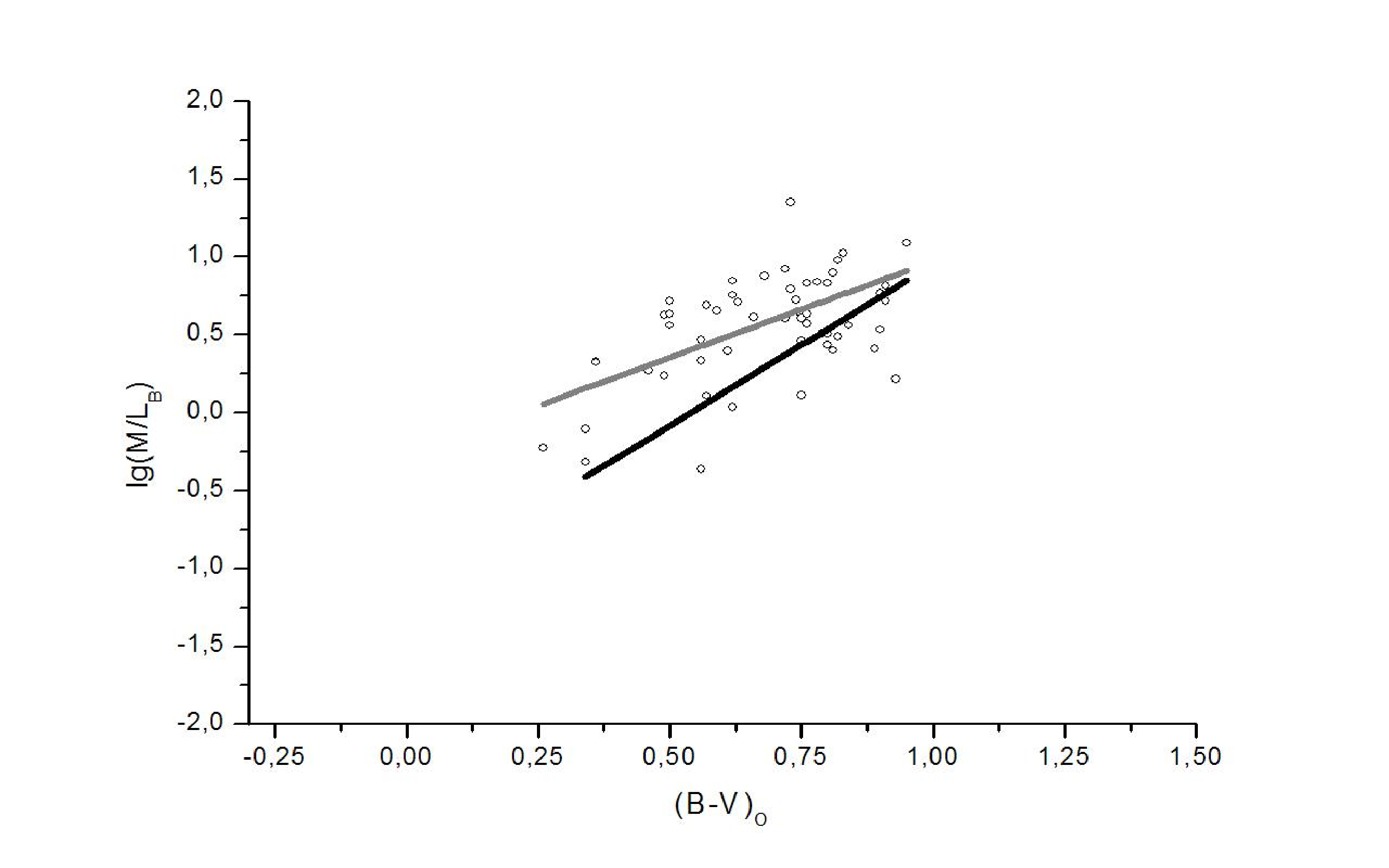}
\caption{Total $M$/$L_B$ ratios within the optical radius $R_{25}$
against the total corrected $(B-V)_0$ colors. a) the entire sample,
b)Virgo galaxies. Gray line shows the least-squares fit (after
2$\sigma$ rejection), black line is a model relationship for
evolving stellar systems (Bell, de Jong, \cite{Bell}).}
\end{figure}

A wide range of $M/L_B$ values, from 30 to 1/30, extending at about
two orders of magnitude, is partially a result of indirect method of
determination of rotation velocity $V$ from the hydrogen linewidth
which in some cases lead to considerable errors, for example, due to
non-circular motions of gas (remember, that $M \sim V^2$).
Nevertheless, for most of the galaxies $M/L_B$ lay in the expected
range between 1 and 10 solar units. Black line in Fig.1 shows the
relation, predicted by the photometric evolution  model (Bell, de
Jong, \cite{Bell}) which uses a modified (bottom-light) Salpeter
$IMF$. The top end of the line corresponds to the galaxies, that
consist of old stars only, whereas the bottom end - to galaxies with
active star formation.

In spite of the wide spread of points on the graph, a bulk of
galaxies follows the well defined sequence, running above the model
relation at some angle to it. After the exclusion of points beyond
$2\sigma$ limits (after two iterations), the correlation is
described by the following linear regression relation:
\begin{equation}
 M/L_B = a_B(B-V)_0 + b_B,\qquad a_B=1.24 ,  b_B=-0.26.
\end{equation}

The difference between the observed and model sequences may be
understood as a result of the presence of a dark halo in galaxies,
which increases $ M/L_B$ ratio. Comparison of the slopes of these
two lines on the graph enables us to conclude that the ratio of
dynamical to stellar masses within $R_{25}$ decreases systematically
from $M/M_* \approx  3-5$ for galaxies with $(B-V)_0 \approx 0.3 -
0.5$ to $\le 1.5$ for the reddest galaxies. This conclusion does not
change if to use $L_V$ luminosity rather than $L_B$. For this case
the coefficients of the linear regression are: $a_V$ = 0.06, $b_V$ =
0.83. Note, that the conclusion about the increasing of the ratio
between stellar and total masses  with $B-R$ color indices was made
by Graham, \cite{Graham} for about a hundred spiral galaxies,
however the correlation between $M/L_B$ and $B-V$ color was not
found by this author.

A similar relation is plotted in Fig. 1b for nearby massive galaxy
cluster Virgo. The  comparison between Figs 1a,b shows that Virgo
galaxies do not differ much in their locations on the diagram.
Despite of different formation conditions, the ratio between the
stellar disc and dark halo masses of cluster galaxies seems to
correlate similarly with color index. It gives evidence that the
relative mass fraction of dark matter within the optical disc is
statistically connected with the stellar composition of galaxies,
that is with the evolution of their stellar population, rather than
with the environment. Dark halo mass contribution appears to be on
the average lower for galaxies in which the fraction of young stars
is low. The relation between the amount of the dark matter and the
star formation history requires a special investigation. It may be
supposed, that it is caused by the influence of large-scale
gravitational instabilities of stellar-gas disc of a galaxy on star
formation. When the dark halo mass fraction is low, the disc is
self-gravitating, that induces the development of the large-scale
instabilities and decreases the time of gas consumption. As a
result, the present time star formation has the low level of
activity.

Note that correlation between the  $M/L_B$ ratios and morphological
type of galaxies is practically absent. The exception are Irr-
galaxies, for which the $M/L_B$ ratios are usually lower, evidently
as the result of the active star formation. We also didn't find a
statistically significant correlation between $M/L_B$ or $M/L_V$
ratios and $HI$ mass, normalized by luminosity (parameter $hic$ in
HyperLeda \cite{leda}), which one would expect, if the dark matter
contribution were connected with the mass fraction of $HI$, as it
was supposed by several authors (see for example, Pfenniger, Revaz,
\cite{HI}). A correlation coefficient between $lg (M/L_B)$ and $hic$
for our sample was found to be as low as 0.2.

It is remarkable that some galaxies have $M/L_B$ ratios, which are
considerably lower than those predicted by the model $M/L$-color
correlation for stellar systems. This disagreement is enhanced if to
take into account the presence of dark halo in these galaxies.
Either the errors of estimations, or the exotic stellar mass
function with a deficit of low massive stars  is required to explain
the observed low $M/L_B$ ratio. This problem will be discussed
further in the next Section.

\section{A comparison of galaxies with high and low $M/L_B$ ratios}

Hereafter we consider in more detail the cases of significant
deviations of points on the $M/L_B$-color diagram from the evolution
model correlation. For this purpose we will compare two groups of
galaxies: those with $M/L_B<1$ and with $M/L_B >10$. We name these
two groups as ``light'' and ``heavy'' galaxies respectively. In
principle,
 if to seek for the galaxies with extremely low or extremely
 high mass fraction of dark halo, there is a good possibility to find them
 in these two groups (those with low dark matter fraction -
among the ``light'' galaxies and those with high dark halo fraction
- among the ``heavy'' ones).

The distributions of ``light'' and ``heavy'' galaxies and the entire
sample of galaxies (5685 objects) by morphological type and
$(B-V)_0$ are compared in Figs. 2a and 2b. As may be inferred from
the histograms, both ``light'' and ``heavy'' groups consist of
galaxies with different morphological types and  colors. However,
``light'' galaxies are on the average bluer, in the comparison with
the general sample (the mean values of $(B-V)_0$ differ by 0.1).
However, some red objects also may be found among the ``light''
galaxies.

\begin{figure}
\includegraphics[width=8cm,keepaspectratio]{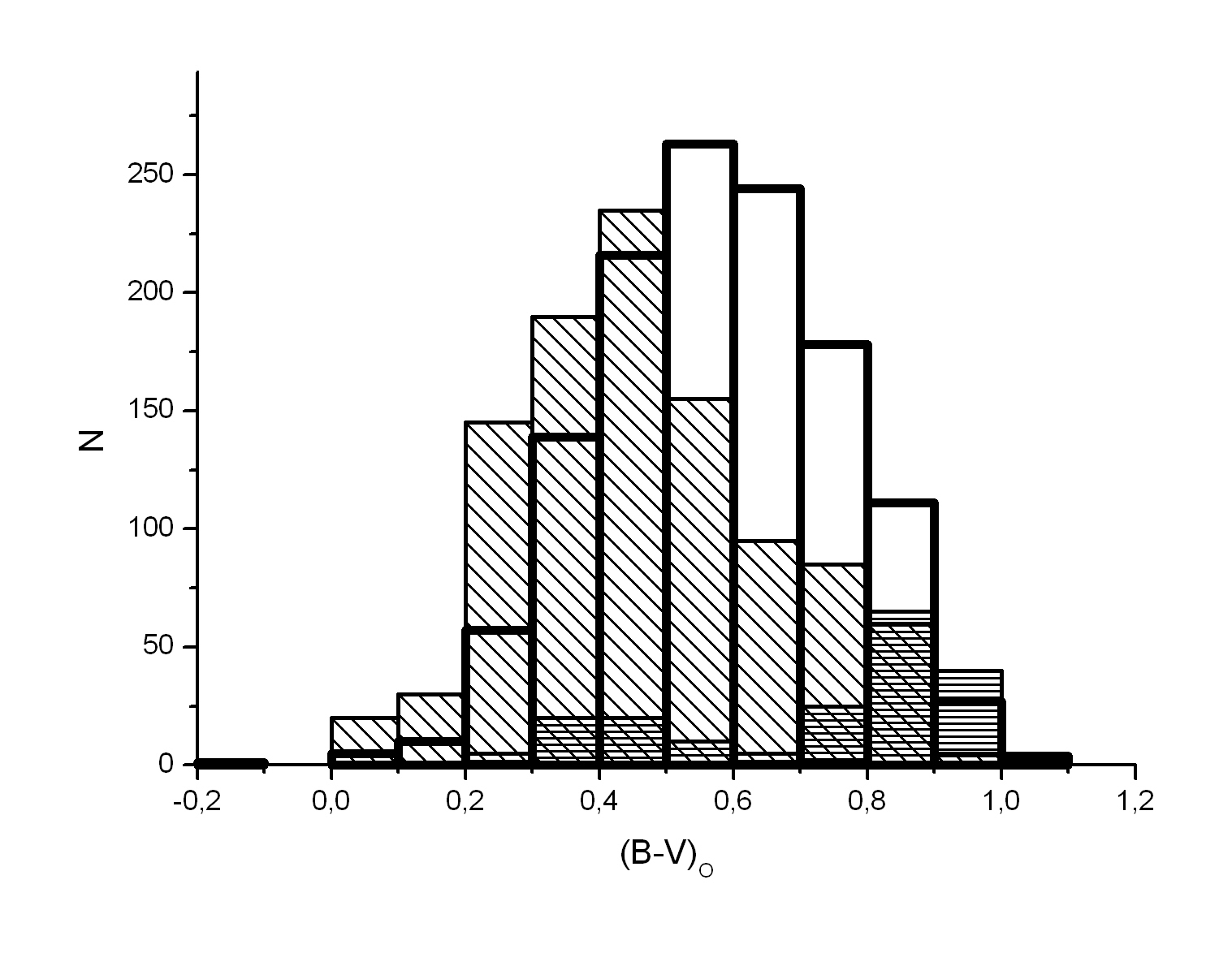}
\includegraphics[width=8cm,keepaspectratio]{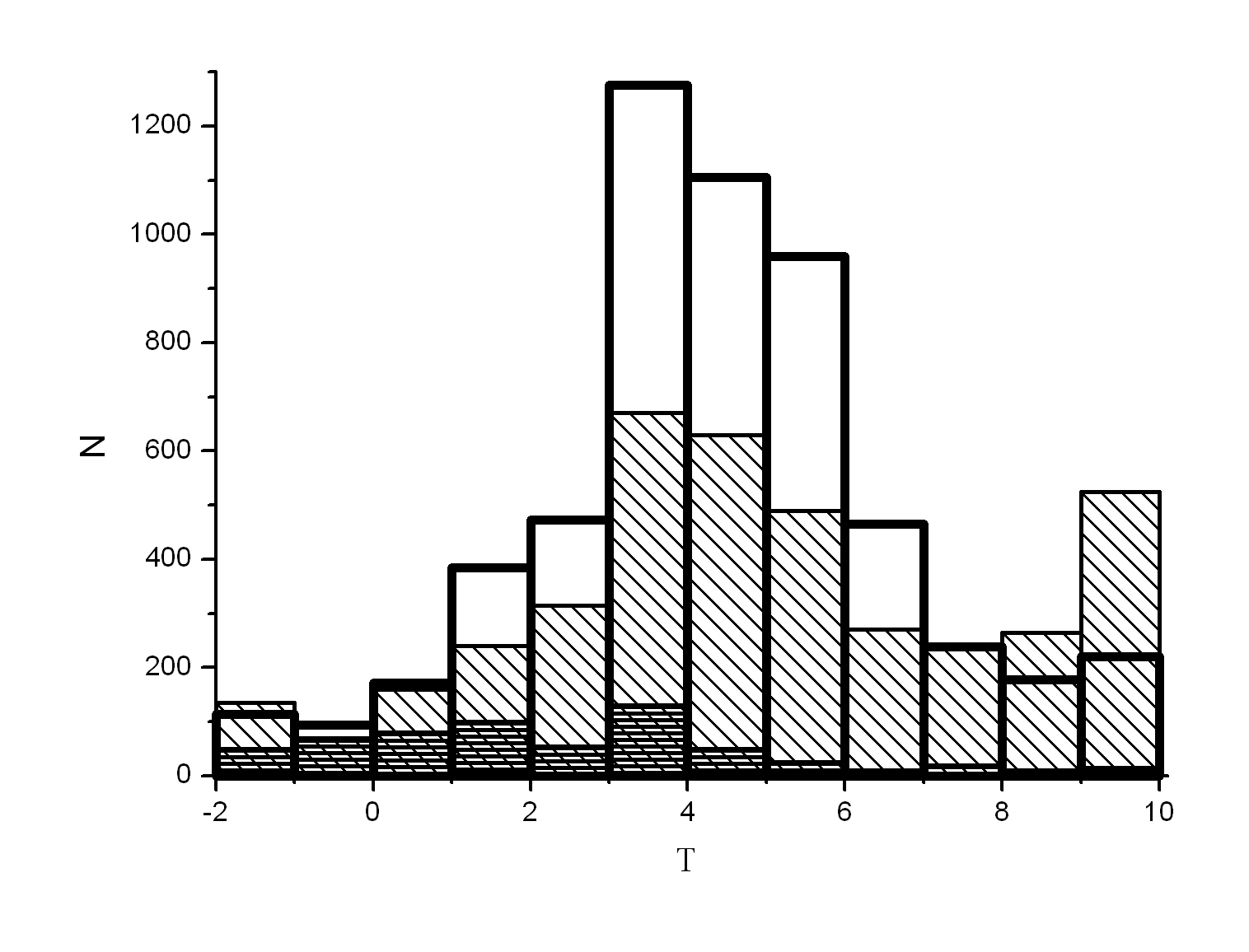}
\caption{Histograms, showing the distribution of galaxies by
$(B-V)_0)$ color (a) and by morphological type (b). The thick line
shows the entire sample of galaxies, cross hatch relates to
``light'' galaxies (($M/L_B<1$), horizontal hatch relates to
``heavy'' galaxies ($M/L_B>10$). The number of ``light'' and
``heavy'' galaxies is multiplied by the factor of 5 for a comparison
convenience.}
\end{figure}

The color of galaxies with $M/L_B>10$ in most cases corresponds to
old stellar population ($(B-V)_0>0.7$). Nevertheless, about one
third of these galaxies demonstrates rather low color index
($(B-V)_0<0.5$). They appear to have on average lower luminosity, so
the conclusion about high mass fraction of dark matter is quite
reasonable for them. In general, galaxies with high $M/L_B$ ratio
are massive systems of S0--Sbc (T= -2 -- +4) types. Situation is
different for galaxies with low $M/L_B$: they have practically the
same frequency of occurrence among various morphological types,
including early types S0--Sa (T=-2 -- +1). About 40\% of ``light''
galaxies have comparatively low dynamical mass $M\le
10^{10}M_\odot$. On the contrary, the mean mass of ``heavy''
galaxies is higher than that for the other galaxies, although their
mean luminosity is practically the same as for the whole sample.

To reduce the probability of significant errors in $M/L_B$
estimations we choose those galaxies which have the  measured
rotation curves, because they allow to find the maximum rotation
velocity more reliably than from the $HI$ linewidth.  Mass $M$ for
these galaxies was also derived from the equation (\ref{formula1}),
where the velocity $V$ was taken from the curve of rotation. The
rotation curve sources were mostly found in the Resolved kinematical
data catalog from HyperLeda database \cite{leda}. If two or more
references are found for the rotation curve of the same galaxy, the
curve with lower scatter of points was favored. More than 50
galaxies from  both samples with published rotation curves were
found, most of them for the ``light'' galaxies.

The comparison of rotation velocities $V$ derived from $HI$
linewidth with those taken from the rotation curve reveals a good
agreement, although there are a few objects with a significant
discrepancy, exceeding 50 km/s, between these estimations. As the
next step, we excluded from further consideration those galaxies,
where the new estimates of $M/L_B$ based on the rotation curves
allowed to exclude them from the ``light'' or ``heavy'' samples.
After this procedure the number of all considered galaxies decreased
roughly by a quarter (roughly by a third for ``heavy'' galaxies).
Some of properties of these galaxies are listed in Tables
\ref{table1} and \ref{table2}. To improve the statistics, the
galaxies with $1\le M/L_B \le 2$ were added to the list of the
``light'' galaxies.
Tables \ref{table1} and \ref{table2} contain:\\
(1) -- Name \\
(2) -- Distance D in Mpc \\
(3) -- Inclination \\
(4) -- Morphological type \\
(5) -- $M/L_B$ ratio \\
(6) -- $R_{max}/R_{25}$ ratio, where $R_{max}$ is the radius,
at which the measured rotation curve extends. \\
(7) -- Source of rotation curve \\
(8) -- Method of rotation velocity measurement (o - optics, r -
radio)\\
(9) -- Notes (g - galaxy in group, c -in cluster, p - in
non-interacting pair, p,i - in interacting pair)

\begin{table}[h!]
\small \caption{``Light'' galaxies with known rotation curves.
\label{table1}}
  \begin{center}
    \begin{tabular}{|c|c|c|c|c|c|c|c|c|}
    \hline
      Galaxy& D, Mpc & $i, ^0$  & t & $M/L_B$ & $R_{max}/R_{25}$ & Source & Estimation & Notes\\
    &&&&&&&method&\\
    (1) & (2) & (3) & (4) & (5) & (6) & (7) & (8) & (9)\\
    \hline
    ESO 008-001&57.3&89&SB(s)d&0.9&0.83&\cite{6}&o&--\\
\hline
ESO 038-012&66.8&68&SAB(rs)bc&1.0&1.25&\cite{6}&o&g\\
\hline
ESO 121-006&13.4&87&Sc&0.5&0.93&\cite{6}&o&--\\
\hline
ESO 338-004&37.7&55&S-?&0.1&0.63&\cite{8}&o&--\\
\hline
ESO 490-028&26&80&Sb&1.0&0.91&\cite{6}&o&p\\
\hline
ESO 565-011&61.1&33&S0-a&1.8&0.95&\cite{9}&o&--\\
\hline
ESO 586-002&88.6&75&S&0.3&1.36&\cite{6}&o&--\\
\hline
NGC 1569&2.2&60&IBm&0.1&1.89&\cite{10}&r&--\\
\hline
NGC 2550&32.1&72.6&Sb&0.7&0.71&\cite{11}&o&--\\
\hline
NGC 3396&21.4&90&Ibm&0.2&0.27&\cite{12}&o&p, i\\
\hline
NGC 3991&43.4&90&IB&0.6&0.95&\cite{13}&o&g\\
\hline
NGC 4668&20.4&67.5&SB(s)d&0.9&0.7&\cite{11}&o&p\\
\hline
NGC 5134&21.8&54.9&SABb&0.5&0.63&\cite{11}&o&p\\
\hline
NGC 4826&7&59&SAab&1.7&2.45&\cite{14}&r&c\\
\hline
NGC 5954&27.2&61&SAB(rs)cd&0.7&1.23&\cite{15}&o&p, i\\
\hline
NGC 6814&20.7&22&SAB(rs)bc&0.3&2.41&\cite{11}&o&--\\
\hline
UGC 03685&26.8&31&SB(rs)b&0.4&4.27&\cite{16}&r&--\\
\hline
ESO 215-039&55.12&48&SABc&0.6&0.95&\cite{6}&o&--\\
\hline
NGC 4618&6.3&35&SBm&1.3&2&\cite{7}&r&p, i\\
\hline
IC 4221&37.33&64.6&SBc&1.7&0.75&\cite{11}&o&--\\
\hline
NGC 1160&35.25&62&SBc&1.9&0.72&\cite{18}&o&p\\
\hline
NGC 3504&20.04&53.4&Sab&1.3&0.78&\cite{7}&r&p\\
\hline
NGC 4424&4.37&61&Sba&1.0&0.2&\cite{19}&o&c\\
\hline
NGC 7743&24.7&40&S0-a&0.4&0.1&\cite{20}&o&p\\
\hline
NGC 4016&49.2&60&Sd&0.5&2.9&\cite{7}&r&p\\
\hline
UGC 11748&74.4&81&Sbc&1.7&1.86&\cite{22}&r&--\\
\hline
NGC 4214&7&38&  I&1.5&5.18&\cite{33}&r&--\\
\hline
NGC 5347&32.50&45&  Sab&0.4&0.21&\cite{34}&o&--\\
\hline
NGC 5850&34.2&37&  Sb&1.4&1.01&\cite{35}&o+r&p\\
\hline
NGC 4490&8.43&45&  SBcd&0.4&1.48&\cite{36}&r&p, i\\
\hline
NGC 2469&47.3&48.3&  Sbc&1.1&1.0&\cite{12}&o&--\\
\hline
NGC 5921&25.2&49.5&  Sbc&0.9&1.65&\cite{37}&o&--\\
\hline
ESO 320-024&37.17&51&  Sc&1.0&0.54&\cite{6}&o&g \\
\hline
NGC 3456&53.67&43&  Sc&1.6&0.57&\cite{6}&o&--\\
\hline
NGC 6012&25.8&46.8&  SBab&1.3&0.94&\cite{34}&o&--\\
\hline
IC 4722&64&45&Sc&1.0&0.80&\cite{6}&o&--\\
\hline
NGC 3359&14.72&53&Sc&1.8&0.98&\cite{40}&o&--\\
\hline
NGC 0864&21.41&47.6&  SABc&1.9&0.90&\cite{12}&o&--\\
\hline
ESO 266-015&39.64&47.8&  Sbc&1.5&0.78&\cite{6}&o&g\\
\hline
NGC 3810&12.17&48.3&Sc&0.6&0.40&\cite{18}&o&--\\
\hline
NGC 7753&71.36&47&  SABb&1.5&1.63&\cite{41}&o&p, i\\
\hline

\hline
    \end{tabular}
  \end{center}
\end{table}

\begin{table}[h!]
\small \caption{``Heavy'' galaxies with known rotation
curves.\label{table2}}
  \begin{center}
    \begin{tabular}{|c|c|c|c|c|c|c|c|c|}
 \hline
       Galaxy & D, Mpc & $i, ^0$  & t & $M/L_B$ & $R_{max}/R_{25}$ & Source & Estimation & Note\\
    &&&&&&&method&\\
    (1) & (2) & (3) & (4) & (5) & (6) & (7) & (8) & (9)\\
    \hline
IC 5063&44.3&50.4&  S0-a&14.2&0.42&\cite{42}&o&--\\
\hline
IC 5096&40.7&90&Sbc&10.1&0.57&\cite{43}&o&--\\
\hline
NGC 0217&53.5&90&S0a&11.4&0.62&\cite{11}&o&--\\
\hline
NGC 2772&42.9&54.1&  Sb&17.8&0.75&\cite{6}&o&--\\
\hline
NGC 4013&11.7&90&  Sb&11.2&2.4&\cite{46}&r&--\\
\hline
NGC 4157&11.2&90&  SABb&10.9&1.7&\cite{47}&r&g\\
\hline
NGC 4772&17&67.5&  Sa&12.9&1.5&\cite{49}&r+o&--\\
\hline
NGC 4866&17&90&S0-a&24.9&1.8&\cite{50}&r&--\\
\hline
NGC 5290&35.5&80.3&Sbc&10.7&0.58&\cite{7}&r&p\\
\hline
NGC 5635&58.4&72.8&  Sb&14.4&0.85&\cite{53}&r&--\\
\hline
NGC 5908&46.1&65.3&  Sb&10.3&0.48&\cite{12}&o&p\\
\hline
UGC 03410&53.9&74.3&Sb&14.9&0.61&\cite{55}&o&p\\
\hline
UGC 12591&95.1&84.6&S0-a&10.7&0.62&\cite{11}&o&--\\

\hline
    \end{tabular}
  \end{center}
\end{table}

From Tables \ref{table1} and \ref{table2} it follows, that the
``light'' and ``heavy'' galaxies exist among
 -- the relatively isolated systems and group members, which
 confirms that the environment does not
affect strongly the total $M/L_B$ ratio.

With the exception of two actively star forming galaxies with poor
quality rotation curves, which possess the lowest values of  $M/L_B
<0.15$ (NGC 1569 and ESO 338-004), the minimum $M/L_B$ ratio in the
galaxies is about 0.3--0.5. Such low values indicate the domination
of young stellar population in their radiation and also the absence
of massive dark halo. Abnormally low  $M/L_B <0.5$ occurs also in
following objects: ESO 586-002 (S), NGC 3996(IBm), UGC 03685 (SBb),
NGC 7743 (S0a), NGC 4016 (Sd), NGC5347(Sab), NGC 4490 (SBcd), and
NGC 6814 (SAB(rs)bc).

The ``heavy'' galaxies with the highest $M/L_B$ ratios are: IC 5063
(S0-a), NGC 2772 (Sb), NGC 4866 (S0-a), NGC 5635(Sb) and UGC 3410
(Sb). All of them have $M/L_B >14$.  However, galaxies with $M/L_B >
20$,  that is at least twice higher than the ratio expected for the
old stellar population, are absent in our sample. The only possible
exception is $NGC~4866$ - an early type spiral galaxy with nearly
edge-on disc. It is worth noting that a significant part of
``heavy'' galaxies in Table \ref{table2} has inclination $i>80^0$,
so at least some of them may fall into this category due to
underestimation of their internal extinction.

 If not to consider the  significant systematical errors of
 evaluations  as the cause of extreme values of $M/L_B$,
 two possible interpretations remains:
 either ``light'' and ``heavy'' galaxies have respectively too light
 and too heavy dark halo, or there exist some
peculiarities of disc stellar population. Extremely low $M/L_B$
ratio may also indicate a burst of star formation in the ``light''
galaxies with relatively low mass fraction of dark halo. The latter
explanation is suitable for 5 galaxies, listed in Table \ref{table1}
(NGC 1569, ESO 338-004, NGC 3991,NGC 1140, NGC 7714 ), since their
blue color  ($(B-V)_0 \le 0.4, (U-B)_0\le -0.4$) shows that young
stellar population gives significant input to their blue luminosity.
However the color indices of other galaxies are considerably redder
with respect to the normal color sequence of galaxies at the
$(B-V)-(U-B)$ diagram.

To reduce the influence of young stars on the $M/L$ evaluation, we
also considered the luminosity of galaxies in the K-band using total
2MASS K-magnitudes, taken from HyperLeda \cite{leda}. Note that in
the case of the low surface brightness galaxies a strong
underestimation of luminosity in the 2MASS $K$-band may be as high
as two stellar magnitudes \cite{Noordermeer}. By this reason NGC
4395 - a galaxy with the lowest surface brightness of our sample
(effective surface brightness in the B-band according to \cite{leda}
is 23.8), was excluded from our consideration.

Fig.3 shows the Tully-Fisher relation for the ``heavy'' and
``light'' galaxies, where the luminosity $L_K$, which characterizes
a stellar population mass, is compared with maximum velocity of
rotation, taken from rotation curve. The straight line represents
the relation for spiral galaxies, given by  Rijke et al.,
\cite{Rijke} using the data presented by Tully and Pierce,
\cite{Tulli}). Although the galaxies we consider were selected on
the basis of $M/L_B$ ratios, Fig. 3 demonstrates that even in the
K-band our ``light'' and ``heavy'' galaxies have respectively higher
or lower luminosity than it is expected from their rotation
velocities, so the discrepancy may exceed an order of magnitude.
This fact rules out the interpretation of ``heavy'' and ``light''
galaxies as a product of errors of light extinction correction,
since  the extinction effect for the $K$ band is much lower than for
the $B$ band. Thereby, as the diagram shows, the galaxies we
consider  are poorly described by the Tully-Fisher relation. If
their distances were determined from this relation, they would be
strongly under- or overestimated.

\begin{figure}
\includegraphics[width=15cm,keepaspectratio]{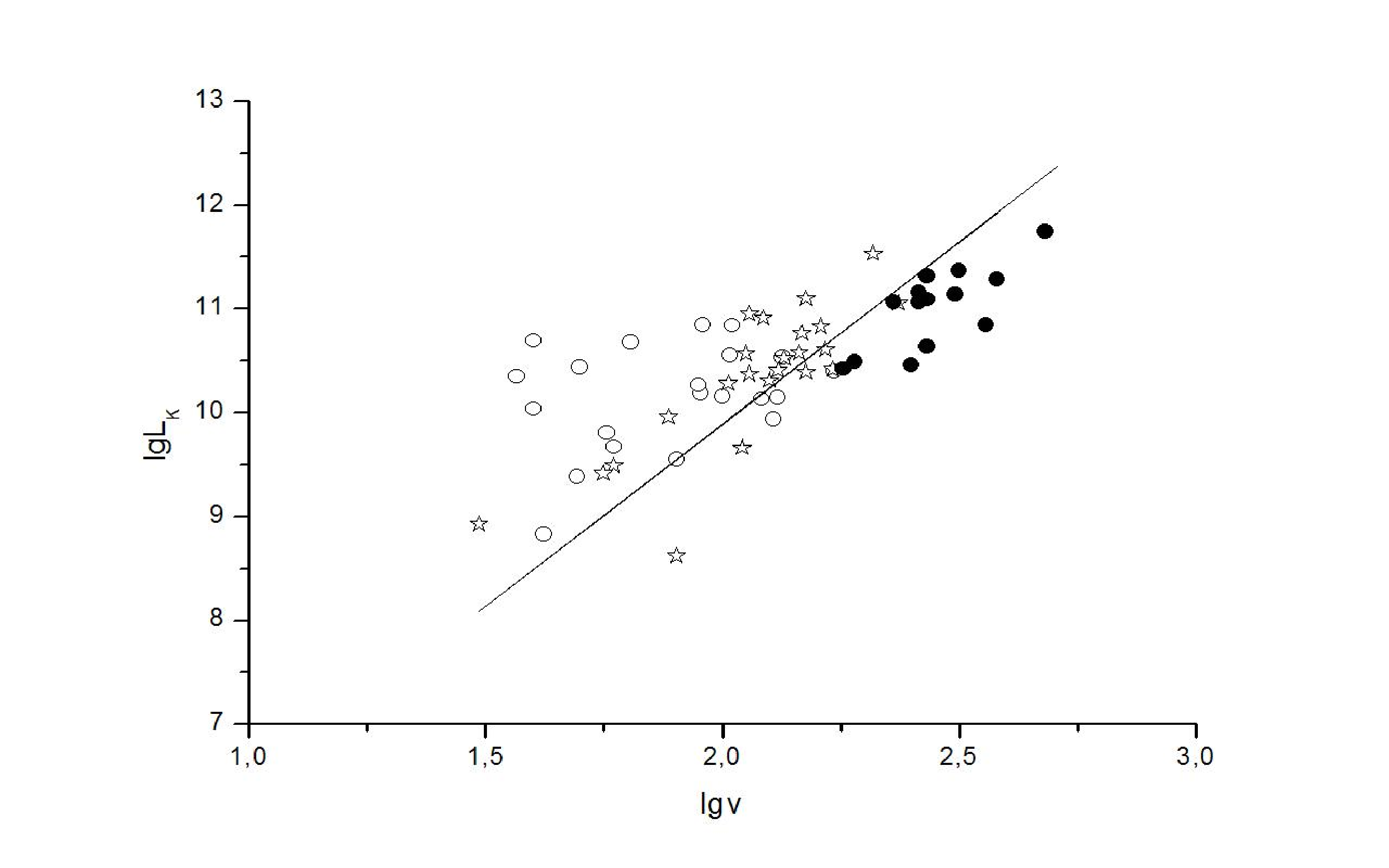}
\caption{The Tully-Fisher relation for the ``light''  (open circles
and asterisks show the objects with $M/L_B<1$ and $1 \le M/L_B \le
2$ respectively), and for the ``heavy'' galaxies (filled circles).}
\end{figure}

To clarify the possible causes of abnormal $M/L_B$ ratio, we carried
out a decomposition of rotation curves, considering stellar disc
$M_*/L$ ratios in each galaxy as a free parameter. It enabled us to
compare this ratios with the model ratios $M*/L$, inferred from the
color of stellar population, for galaxies  with reliable rotation
curves and color indices found from the literature. The rotation
curves were fitted by a simple model, that consists of three
components: King's bulge, thin exponential disc and
pseudo-isothermal halo. The resulting estimation of the disc mass is
poorly sensitive  to the choice of the bulge parameters, but it
depends strongly on the adopted disc radial scalelenght which is
assumed to be close to the photometric radial scalelength. In the
cases when the disc surface brightness distribution is poorly fitted
by exponential law, we used the observed photometric profile
assuming a constant mass-to-light ratio (for such galaxies there is
a dash in the column "Disc central surface density" in Tables
\ref{table3}, \ref{table4} ). A well-known ambiguity of rotation
curve decomposition forced us to employ the maximum disc model. This
model generally does not lead to significant overestimation of disc
mass, unless slowly rotating galaxies are considered (see, for
example Kranz et al., \cite{Kranz}).

Some examples of our rotation curve decomposition are shown in Figs.
4a, b. The results of the  modeling of
the ``light'' and ``heavy'' galaxies may be found in Tables \ref{table3},\ref{table4}.
Their columns show:\\
(2)-- Mass of the disc within $R_{25}$, \\
(3)-- Total to disc mass ratio within $R_{25}$,\\
(4), (5)--  $M_d/L_B$ and $M_d/L_K$ ratios for the disc component,\\
(6)-- $(B-V)_0$ color,\\
(7)-- Central disc density, extrapolated to $R=0$,\\
(8)-- The reference sources of the disc radial scalelengths,that
were used in the modeling.

\begin{figure}[!h]
\includegraphics[width=8cm,keepaspectratio]{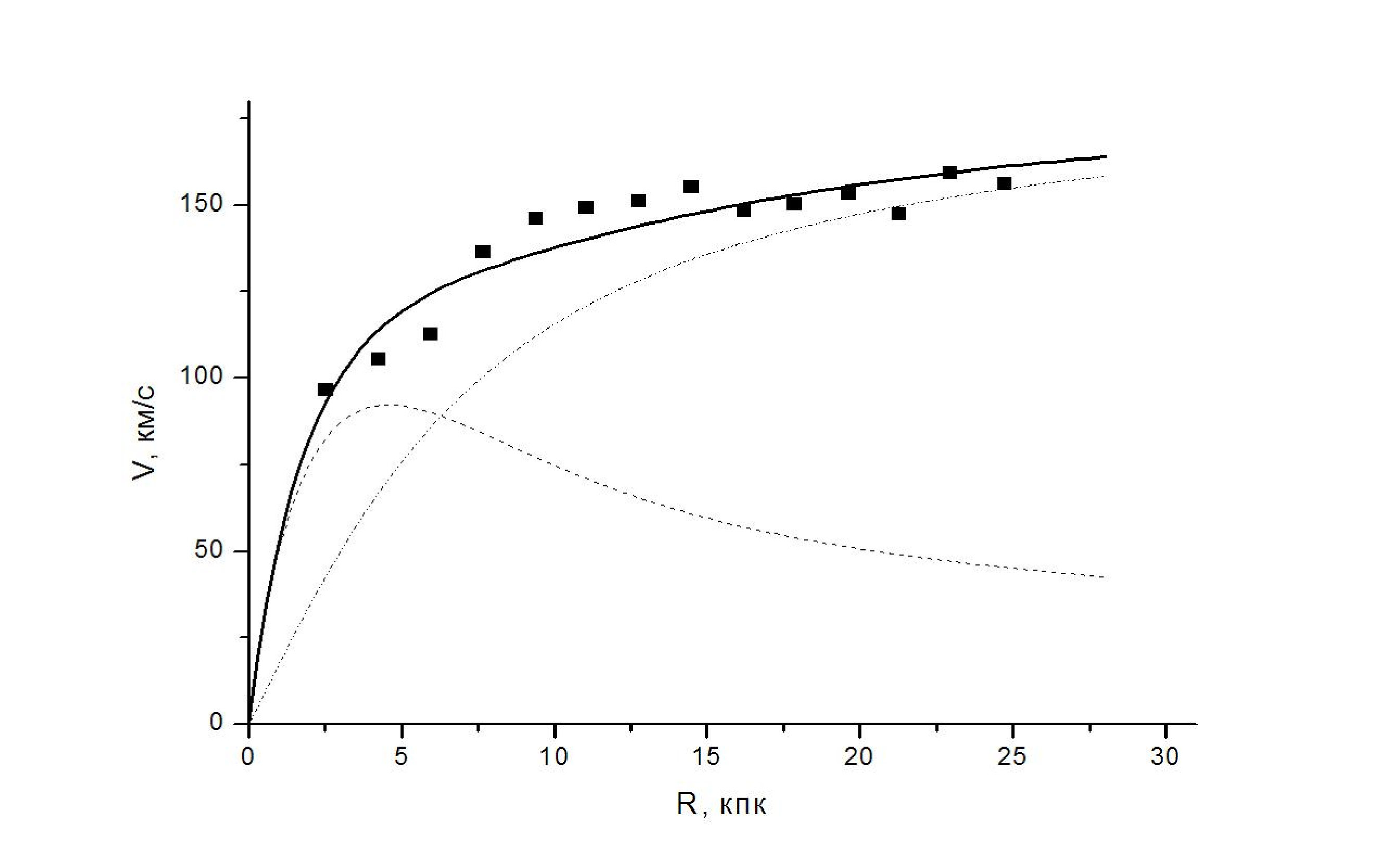}
\includegraphics[width=8cm,keepaspectratio]{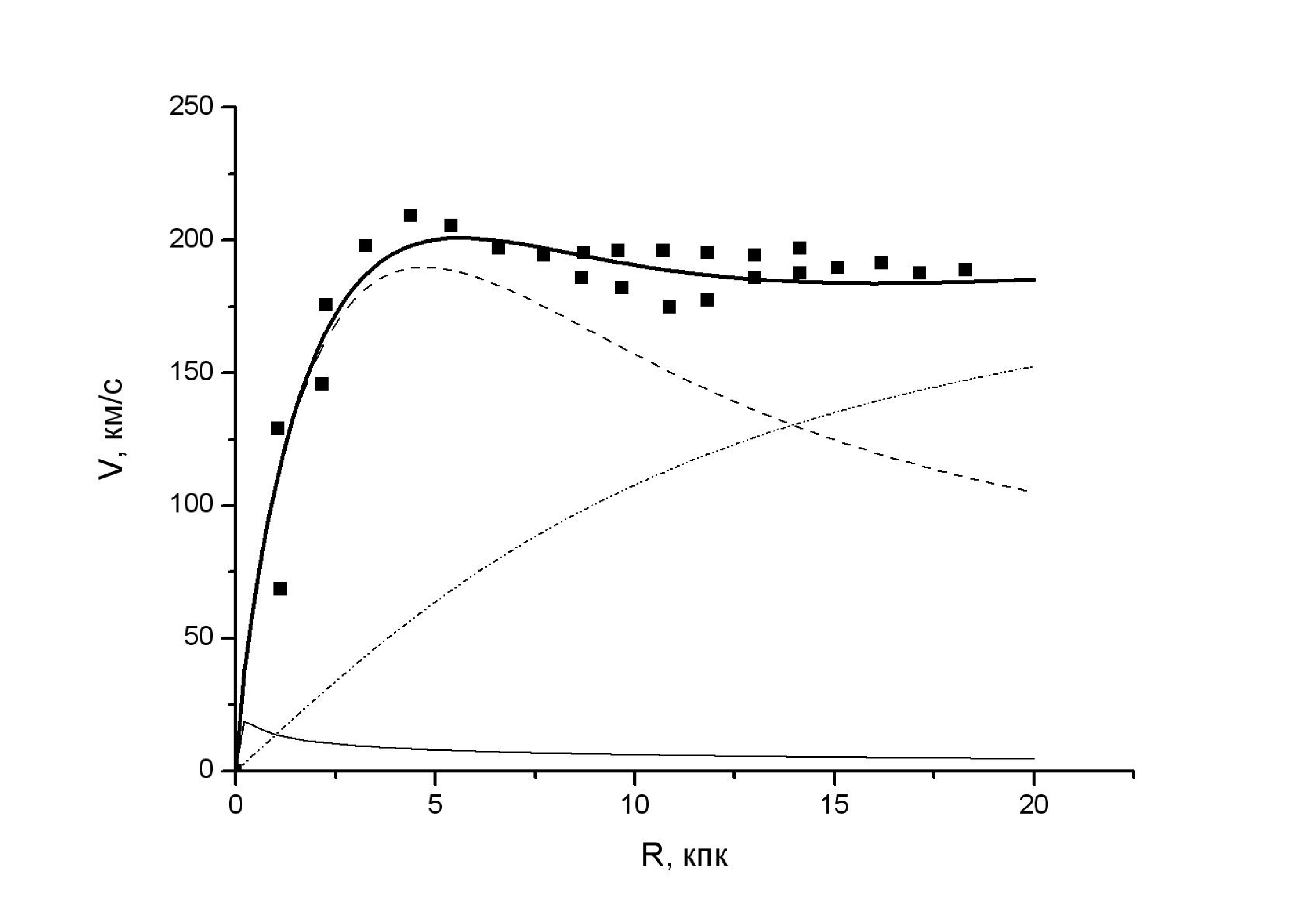}
\caption{The examples of rotation curve decomposition. a) $NGC~4826$
(a ``light'' galaxy), b) $NGC~ 4157$ (a ``heavy'' galaxy). }
\end{figure}

\begin{table}[p]
\small \caption{Galaxies with $M/L_B<$2: the results of modeling.
\label{table3}}
  \begin{center}
    \begin{tabular}{|c|c|c|c|c|c|c|c|}
 \hline
  Galaxy& $M_{tot}/M_\odot$ & $M_{disc}$ & $M_{disc}/L_B$ & $M_{disc}/L_K$ & $(B-V)_0$ & $\Sigma_0$ & Photometry\\
 (1)&(2)&(3)&(4)&(5)&(6)&(7)&(8)\\
 \hline
ESO008-001&2.79E10&0.99&0.87&1.72&-&-&\cite{6}\\
\hline
ESO 038-012&7.34E10&0.36&0.37&1.58&-&-&\cite{6}\\
\hline
ESO 121-006&3.26E10&0.24&0.13&3.31&0.62&-&\cite{6}\\
\hline
ESO 490-028&4.19E9&0.21&0.21&1.20&-&-&\cite{6}\\
\hline
ESO 565-011&4.43E10&0.44&0.81&0.47&0.73&-&\cite{9}\\
\hline
ESO 586-002&2.07E10&0.95&0.32&-&-&-&\cite{6}\\
\hline
NGC 2550&1.7E10&0.45&0.34&1.11&-&390&-\\
\hline
NGC 3396&3.76E9&1.0&0.23&0.31&-&84&-\\
\hline
NGC 3991&2.5E10&0.45&0.83&-&0.36&320&-\\
\hline
NGC 4668&4E9&0.24&0.23&0.76&0.38&95&\cite{27}\\
\hline
NGC 5134&7.3E9&0.84&0.42&0.13&0.66&220&\cite{27}\\
\hline
NGC 4826&4.25E10&0.25&0.43&0.55&0.71&440&\cite{27}\\
\hline
NGC 5954&1.26E10&0.57&0.40&0.79&-&-&\cite{15}\\
\hline
NGC 6814&1.36E10&0.62&0.18&0.17&0.67&240&\cite{28}\\
\hline
UGC 03685&3.99E9&0.19&0.06&0.16&-&12&-\\
\hline
ESO 215-039&2.71E10&1.0&0.63&0.70&0.31&-&\cite{6}\\
\hline
NGC 4618&3.39E9&0.45&0.58&1.16&0.38&250&\cite{28}\\
\hline
NGC 5850&6.63E10&1.0&1.33&0.46&0.72&339&\cite{28}\\
\hline
NGC 2469&2.25E10&0.45&0.28&0.79&-&600&-\\
\hline
NGC 5921&1.1E10&0.41&0.15&0.14&0.6&77&\cite{28}\\
\hline
NGC 3359&3.69E10&0.54&0.70&1.31&0.4&360&\cite{28}\\
\hline

   \end{tabular}
  \end{center}
\end{table}

\begin{table}[p]
\small \caption{Galaxies with $M/L_B>$10: the results of modeling.
\label{table4}}
  \begin{center}
    \begin{tabular}{|c|c|c|c|c|c|c|c|}
 \hline
  Galaxy& $M_{tot}/M_\odot$ & $M_{disc}/M_{tot}$ & $M_{disc}/L_B$ & $M_{disc}/L_K$ & $(B-V)_0$ & $\Sigma_0$ & Photometry\\
 (1)&(2)&(3)&(4)&(5)&(6)&(7)&(8)\\
 \hline
 IC 5063&3.5E11&0.37&3.9&0.80&0.91&1360&\cite{56}\\
\hline
IC 5096&4.1E11&0.15&3.1&0.70&--&--&\cite{57}\\
\hline
NGC 0217&4E11&0.23&1.9&0.34&--&1150&--\\
\hline
NGC 0936&1.8E11&0.69&12.5&2.02&0.9&2130&\cite{58}\\
\hline
NGC 2772&1.1E11&0.60&2.5&0.62&--&980&--\\
\hline
NGC 4157&7.7E10&0.58&6.3&1.55&0.64&1550&\cite{27}\\
\hline
NGC 4772&1.1E11&0.69&6.2&2.25&0.83&1280&\cite{27}\\
\hline
NGC 5290&2.4E11&0.46&4.0&0.94&-&632&\cite{59}\\
\hline
NGC 5635&6.5E11&0.49&6.0&1.71&--&1680&--\\
\hline
NGC 5908&4.2E11&0.40&4.3&0.94&0.81&1670&--\\
\hline
UGC 03410&2.1E11&0.56&6.9&1.12&--&995&--\\
\hline
UGC 12591&1.1E12&0.45&4.8&0.90&--&5700&--\\
\hline

   \end{tabular}
  \end{center}
\end{table}

The results of the decomposition of the rotation curves  show that
galaxies with different -- high or low -- $M/L_B$ values may have
different ratios of disc-to-total masses. For both ``light'' or
``heavy'' galaxies the disc mass fraction lays in the range 0.4 --
0.7, although a dispersion of this ratio is higher for the ``light''
galaxies. At least partially, a high dispersion may be explained  by
their lower mean luminosity, because  the maximum disc model we used
can significantly overestimate the mass of some low mass galaxies.

The dynamically estimated masses of  the galactic discs $M_d$ allow
us to compare the positions of the ``light'' and ``heavy'' galaxies
on the baryonic Tully-Fisher relation connecting $M_d$ with the
inclination-corrected $HI$ linewidth, which is equal to double value
of rotation velocity (Fig.5). The solid line reproduces the
relation, given by Gurovich et al., \cite{Gurovich} for two samples
of galaxies of normal and low luminosity for which the disc masses
were estimated from their luminosity in the $I$ band. The
contribution of the gas to the disk mass was also taken into account
for slowly rotating systems. With an exception of a few most slowly
rotating galaxies, ``light'' and ``heavy'' galaxies we discuss
follow the general sequence,
 in contrast with the Tully-Fisher relation $L_K - V$ where they deviate
 significantly from the expected relationship (Fig. 3).
 With some exceptions, the masses $M_d$ of the galaxies we consider
 are typical for their velocities of rotation,
 that agrees with the conclusion made above
  that the cause  of their low or high  $M/L$ is  not an unusually low
  or high fraction of dark
halo mass, but rather are the properties of their stellar
population.

\begin{figure}
\includegraphics[width=15cm,keepaspectratio]{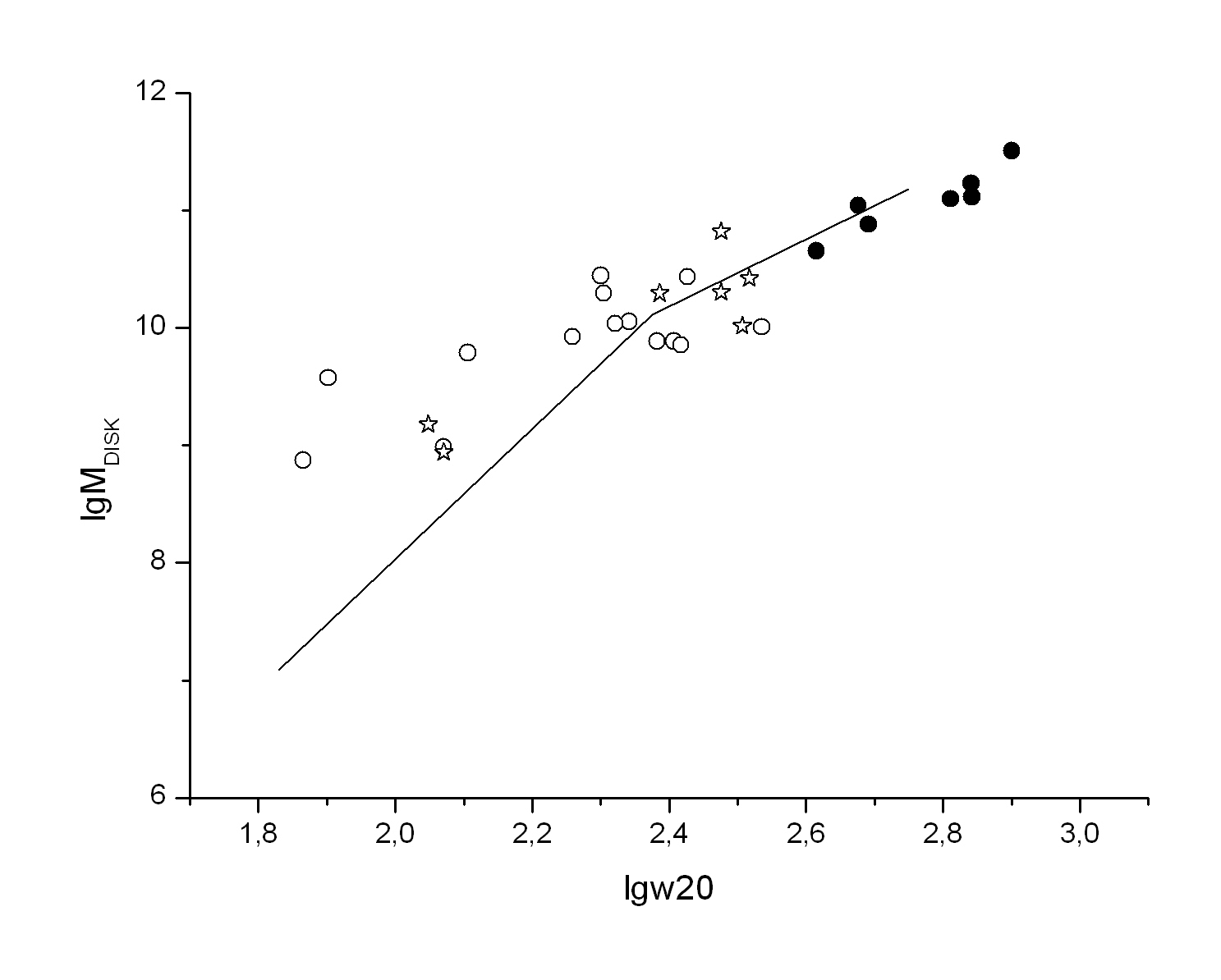}
\caption{The baryonic Tully-Fisher relation $lgM_*-lgW_{20}$ for
``light'' (open symbols) and ``heavy'' (filled symbols) galaxies;
circles and asterisks show galaxies with $M/L_B<1$ and with $1 \le
M/L_B \le 2$ respectively. The solid line reproduces the relation
given by Gurovich et al., \cite{Gurovich}.}
\end{figure}

Fig. 6 shows the $M_*/L_B$-color diagram for the galaxies we
discuss.  The ``heavy'' and ``light'' galaxies are marked by filled
and open symbols, respectively. Galaxies, whose disc masses were
estimated by decomposition of their rotation curves (triangles) and
those where only the  total mass and hence the total $M/L_B$
estimations were possible due to poor rotation curves (squares for
``heavy'' galaxies and circles or asterisks for ``light'' ones), are
plotted at the same diagram. Open asterisks and circles relate to
galaxies with $1 \le M/L_B \le 2$ and with $M/L_B<1$ respectively.
As far as the total $M/L_B$ ratio is the upper limit of disc
$M_d/L_B$ ratio, the corresponding symbols are marked by the arrows.
The vertical lines connect the values of $M/L_B$ and $M_d/L_B$ for
the same object. Here we ignore the difference between the disc mass
$M_d$  and mass of the stellar population $M_*$. The straight line
represents the model relation ``color - $M_*/L$",  for evolution
models of stellar population with the scaled-down (light) Salpeter
$IMF$ (Bell, de Jong, \cite{Bell}). The other shapes of stellar
$IMF$s, suggested by different authors, shift the model sequence
vertically in such a way that the values of log($M_*/L_B$) changes
only within a few tenths of dex (see, for example Portinari et al.,
\cite{Portinari}).

\begin{figure} [h!]
\includegraphics[width=15cm,keepaspectratio]{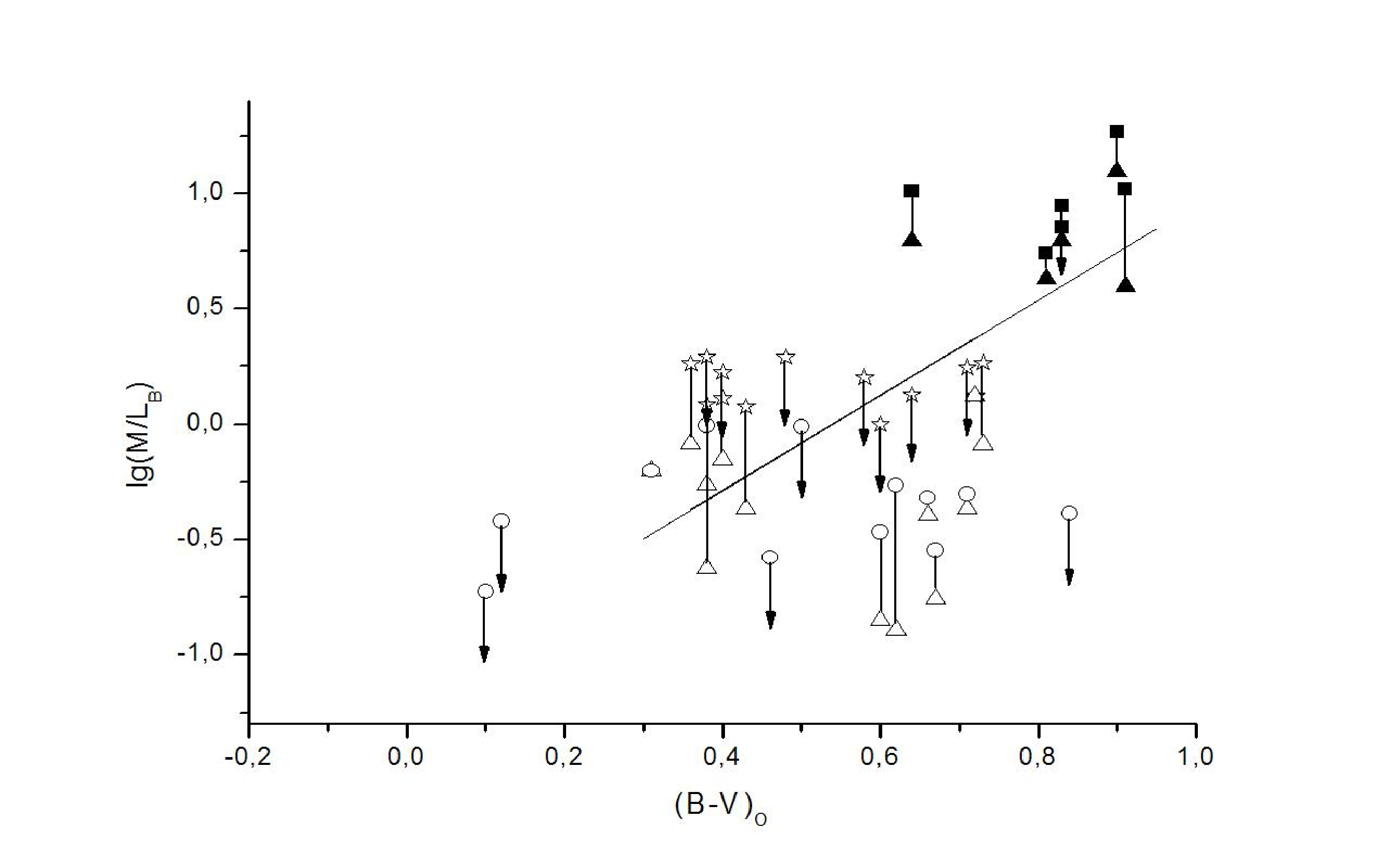}
\caption{The $M_*/L_B$-color diagram for the ``light'' (open
symbols) and ``heavy'' (filled symbols) galaxies. The upper symbols
(squares, circles and asterisks) show the total $M/L_B$ ratios
(circles and asterisks correspond  to the galaxies with $M/L_B<1$
and $1 \le M/L_B \le 2$ respectively). The triangles mark the disc
$M_*/L_B$ ratios. The vertical lines connect the symbols,
corresponding to the  $M/L_B$ and $M_*/L_B$ in the same object. The
solid line represents the model relation for stellar systems
\cite{Bell}.}
\end{figure}

 Fig. 6 shows that the stellar disc $M_*/L_B$ ratios poorly follow
 the expected model sequence. However the
situation is different for the ``light'' and ``heavy'' galaxies. The
$M_*/L$ values for the ``heavy'' galaxies lay systematically above
the model relation, but generally not far from it, especially if to
remember that we used the maximal disc solution. A small number of
``heavy'' galaxies considered here does not allow to make a
confident conclusion about any systematic peculiarities of their
properties. Two cases, however, deserve a special attention. This
first one is $IC 5096$, the galaxy with the highest $M/L_B$. Its
position on the diagram allows to conclude that  there is a
significant predominance of the dark halo mass over disc mass in
this galaxy. The seconds case ($NGC~ 936$, the leftmost of the
filled symbols) suggests a stellar disc, not halo, that is unusually
massive for the observed luminosity and color, indicating the
possibility of exotic stellar mass function.

The situation with the ``light'' galaxies is more interesting. Two
galaxies with the lowest values of $(B-V)_0$ experience a burst of
star formation ($NGC~1569$ and $ESO 338-004$), which explains their
low total $M/L_B$ ratios. The color indices of these galaxies cannot
be described by the simple model \cite{Bell} of stellar systems with
monotonous  evolution of star formation. Most of the other ``light''
galaxies lay below the expected model relation, especially if to
take into account  that the maximum disc model we used may
overestimate $M_*/L$ ratio. It corresponds to the following
galaxies: $NGC~4214, ~5134,~ 5347,~ 5921,~ 6814,~ 7743$ and
$ESO~121-006$. These objects require individual and more detailed
investigation. Their low $M/L_B$ and $M_*/L_B$ ratios cannot be
explained by the excess of OB-stars, since their color $(B-V)_0>0$,
except $NGC~ 4214$. This conclusion remains to be the same with
$M_*/L_K$ instead of $M_*/L_B$ at the diagram.

Discs of the ``light'' galaxies listed above are distinguished not
only by their low mass-to-light ratio, but also by the low surface
density. Their central surface densities obtained from the rotation
curve decomposition lay in the range from 100 (or even less) to a
few hundred $M_\odot / pc^2$ (see Table \ref{table3}). For
reference: the typical central surface brightness of the discs of
HSB spiral galaxies is $\mu_{0B} = 21-22 ^m/sq.arcsec$, hence, for
the typical ratio $M_*/L_B$ = 4 it corresponds to the central
surface density of 400 -- 1000 $M_\odot/pc^2$. Non-typical low
density of a disc may indicate a strong deficit of stars with masses
less than a few solar masses. Their absence affects strongly the
total mass rather than the total luminosity of a disc. Since the
masses of stellar discs are provided mostly by the old stellar
population, the expected bottom-light initial mass function in the
``light'' galaxies may be associated with the early epoch of disc
formation, and not necessary with the present time star formation.

To summarize, there is no sole explanation for the extremely low and
extremely high values of total $M/L_B$ ratios in the galaxies we
consider in this paper. In some cases they are clearly associated
with errors of velocity or luminosity estimations, but in other
cases we encounter with the real features of galaxies, such as
non-typical disc-to-halo mass ratios, or, especially for the
``light'' galaxies,  with stellar population peculiarities such as
burst of star formation or non-typical stellar initial mass
function.

\section{Conclusions}
\begin{enumerate}
\item From the $M/L_B$ -- $(B-V)_0$ diagram, plotted for about 1300
discy galaxies,  it follows that the $M/L_B$ ratio of galaxies
increases  with the color index, although not so steep as it is
predicted by the stellar population evolution models. This allows to
conclude that the ratio between the dark halo and stellar disc
masses decreases along the color sequence towards the ``red''
galaxies with old stellar population. However, some galaxies do not
follow this general trend, and this cannot be explained in all cases
by the errors in their rotation velocity or luminosity estimations.
Virgo cluster galaxies do not outlay by their positions on the
diagram from the others.

\item Extremely high $M/L_B$ ratios $M/L_B >10$ are rare and usually occur
in the ``red''  galaxies. With few exceptions (for example - $IC~
5096$) they possess not too massive dark halos, being normal
galaxies with old stellar population. Nevertheless, there also exist
some `` heavy'' galaxies with a  mixture of old and young stars, so
their high $M/L_B$ ratio can be associated with a very massive dark
halo.

\item   Low  ratios $M/L_B <1$ of galaxies are also not
caused by a sole reason. In some cases they reveal the actively
ongoing star formation, but this interpretation is definitely not
valid for some galaxies: they have an extremely low both total
$M/L_B$ and disc $M_d/L_B$ ratios. These ratios are too low to be in
agreement with the observed color of galaxies even in the absence of
the  halos, indicating the abnormal initial stellar mass function
with a strong deficit of low massive stars.
\end{enumerate}

\section*{Acknowledgments}
We acknowledge the possibility to use the HyperLeda database
(http://leda.univ-lyon1.fr/).

This work was supported by Russian Foundation for Basic Research,
grant 07-02-00792.

\end{document}